# SkyQuery: A Web Service Approach to Federate Databases

Tanu Malik        Alex S. Szalay        Tamas Budavari        Ani R. Thakar

Johns Hopkins University
Baltimore
USA

tmalik@cs.jhu.edu, szalay@pha.jhu.edu, budavari@pha.jhu.edu, thakar@pha.jhu.edu

## Abstract

Traditional science searched for new objects and phenomena that led to discoveries. Tomorrow's science will combine together the large pool of information in scientific archives and make discoveries. Scientists are currently keen to federate together the existing scientific databases. The major challenge in building a federation of these autonomous and heterogeneous databases is system integration. Ineffective integration will result in defunct federations and under utilized scientific data.

Astronomy, in particular, has many autonomous archives spread over the Internet. It is now seeking to federate these, with minimal effort, into a Virtual Observatory that will solve complex distributed computing tasks such as answering federated spatial join queries.

In this paper, we present SkyQuery, a successful prototype of an evolving federation of astronomy archives. It interoperates using the emerging Web services standard. We describe the SkyQuery architecture and show how it efficiently evaluates a probabilistic federated spatial join query.

## 1. Introduction

Autonomous astronomy archives operating over the Internet feel an increasing need to federate with each other. A federation will enable their users to combine data in different archives for the same astronomical object and thus examine each object in more detail. Similar needs to federate are also felt in pharmaceutical [UFI01], medical [Daw00], geographical [Arn98], and other scientific archives.

A federated database is a collection of cooperating but autonomous component archives behaving like a single integrated database [She90]. The challenge in federation arises from the heterogeneity of the autonomous archives. Each archive makes independent choices in its hardware, software, data schema, etc., and invests its resources accordingly. As a result, differences arise among archives in the types of network systems, operating systems, database systems, programming platforms, and algorithmic techniques used.

An archive is naturally reluctant to forsake its autonomy --- even if it isn't, complete restructuring can be prohibitively expensive. Federated database architecture is designed to maintain autonomy and yet accomplish federated tasks. This implies that the interfaces of the architecture must conform to some common standard for interoperability. Such a standard specifies, primarily, a communication protocol and a data exchange format. To be functional, a federation on the Internet needs a communication protocol that is universally acceptable and a data exchange format that is simple and extensible.

In this paper, we present SkyQuery [Sky02], a successful prototype of one of the first proposed federations of astronomy archives. It supports a complex federated query called the cross match query, which is a spatial join query that matches objects between archives, if they correspond to the same astronomical body. SkyQuery is implemented using the emerging middleware standard called Web services [Sim01, XML02]. Its success in supporting cross matches has implications on the feasibility of adopting Web Services as a standard for a full-fledged federation of astronomy archives that also supports a variety of other federated queries and transactions.

The rest of the paper is organized as follows. In Section 2, we give an overview of astronomy archives on the Internet, mentioning characteristics of astronomical data and the reasons behind the need to federate. A lot of effort, including from the geographical information systems community, has been directed towards standards for interoperability [Ama96, Arn98]. Some of the standards that have taken advantage from this community are CORBA [Obj02], DCOM [Dis98], and Java RMI [Jav98]. In Section 3, briefly look at the interoperability issues relevant to Internet federations. We also describe Web Services and point out its advantages and disadvantages. Spatial query execution in distributed and federated environments has recently become a focus of research. In Section 4, we give an overview of the existing work in this area. Section 5 is the heart of the paper. We first describe in detail the architecture of SkyQuery. We then define a cross match query. Next, we present an original algorithm designed to efficiently solve this query. Finally, we present the steps followed by the various components of SkyQuery in the execution of this algorithm. In Section 6 we make concluding remarks.

In this paper, we use the terms "archive" and "database" interchangeably. Also, when we say, "user query", we mean, the cross match query specified by the user.

## 2. Astronomy and the DataAvalanche: the Need for Federation

Astronomy is a particularly good domain for building federations and evaluating techniques for federated spatial queries [APS99]. Astronomers have conducted more than a 100 independent surveys [Slo02, NAS02], each mapping the sky in different wavelengths, but related to each other by the unique position of each astronomical body. Typically, such surveys cover 10 – 100 million objects, with some of the larger ones expected to cover a billion objects. The archives resulting from these surveys are autonomous and heterogeneous. In spite of this, building a federation of these archives has been a goal of astronomers for a long time as the users will then be able to run federated spatial join queries. Through these queries, they would observe the same sky in other wavelengths (using someone else's data) and combine the available observations into a multi-spectral data set. This would immensely aid in making discoveries faster and easier [Sch00].

Astronomical data sets are large and contain a wide-variety of data types—from raw pixels to fully derived tables of scientific measurements [Sch00]. In addition this data is freely available. By studying and providing for the needs of the world's 10,000 strong astronomy community, one can gain valuable insight in developing federated systems – which can then be scaled to support a much wider audience.

## 3. Choice of a Standard for Interoperability

The components of a federated database need to agree upon a flexible communication standard for supporting message passing and for the RPC semantics used between them. To achieve this flexibility, we can either use proprietary standards or existing distributed object frameworks like DCOM, CORBA, Java RMI, etc., which have provided excellent solutions for distributed and multi-databases [Ozs99, Ama96, Arn98]. Proprietary solutions provide limited flexibility, as they are usually tailored to satisfy the needs of a single enterprise where the kind of heterogeneity is not the same as witnessed over the Internet. Similarly, distributed object frameworks for multi-databases are poorly suited to federated database systems in the following ways:

1. <u>Network protocol dependence</u>: These standards are integrated closely with a communication protocol. For example, CORBA 2.0 compliance requires vendors to support the TCP/IP protocol suite.
2. <u>Operating system dependence</u>: Most of the existing standards have platform preferences. DCOM is tightly integrated with Windows while CORBA with UNIX-based platforms. For DCOM, porting the solution can be arduous; while in CORBA, it is not guaranteed that ORBs [Obj02] from different vendors are interoperable.
3. <u>Semantic dependence</u>: Standards like DCOM, and Java RMI are based on the object-oriented paradigm. Implementing them is easy only in an object-oriented language.
4. <u>Unwieldy</u>: These standards are reliable, secure, and scalable, but at the same time, overloaded with functionality. This implies that solutions based on them are complex and difficult to maintain, and so unsuitable for applications not making use of the extra functionality they provide.

In the following subsection, we give a brief description of the emerging Web services standard.

### 3.1 Web Services

A distributed computing model consists of a message exchange model, a communication protocol, and mechanisms for describing, defining, and discovering services. Web Services use Internet-based application-level protocols like Simple Mail Transfer Protocol (SMTP) [Smt01] and Hypertext Transfer Protocol (HTTP) [Hyp99] for communication between applications. Simple Object Access Protocol (SOAP) [Sim01] is the message-exchange protocol; Web Services Description Language (WSDL) [Web01] is the standard for describing and defining, services; and Universal

Description, Discovery and Integration (UDDI) [Uni02] is the standard for discovering services.

SOAP is a simple and lightweight mechanism for exchanging structured and typed information between peers in a decentralized, distributed environment. SOAP supports both one-way and request-response message exchange paradigms. Further, it is a flexible, easy-to-deploy, loosely coupled, access-protocol-independent, open standard protocol. It is highly flexible as it uses XML [Ext96] as the data exchange format. The syntax-independence of XML allows communicating entities to specify the syntax of the data (meta-data) along with the actual payload, making it highly flexible for an evolving payload. SOAP message processing entities are lightweight, as it doesn't define bulky protocol-dependent headers. HTTP messages containing SOAP need to specify only one extra field "Soap Action", below custom HTTP headers, that identifies the SOAP entity on the destination machine. SOAP is easy-to-deploy as it currently uses XML and HTTP, which are ubiquitous easy-to-understand de facto Internet standards. Almost all machines support the HTTP protocol, and XML parsers are freely available, and being text based, very easy to understand and debug. SOAP is loosely coupled, as, due to the flexibility of XML, it doesn't rely on any programming model or application semantics. SOAP is also access-protocol-independent.WSDL consists of two distinct parts - service definition and service implementation. Service definition is an XML-style description of what the service intends to provide, i.e., names of messages and their parameters, type of messages, etc. Service implementation specifies binding to a particular protocol or data type, i.e., syntax of the messages exchanged, protocols used to transfer messages, etc. By dissociating service definition from its implementation, WSDL allows re-use of the service description interface by clients that might be using other programming models to implement the service. Currently WSDL supports only SOAP and HTTP protocols for message communication.

Services need a unique service for discovering other services that may be available on the Internet. This service should provide a common repository where services can register themselves and be discovered. UDDI is the standard architecture for building such repositories.

To summarize, the simplicity and flexibility of XML and the ubiquity of HTTP make Web Services an attractive solution for federating archives.

## 4. Related Work

Federated database systems have been extensively researched [Lit90]. Most studies, however, have concentrated on schema integration, cost estimation for sub-query execution at individual archives, construction of query execution plans, transaction management, and other issues that are not the primary focus of this paper and were only partially considered while building SkyQuery.

Spatial joins, which are much more expensive than regular relational joins, have also received a lot of research attention [Arg98, Bri93]. But most schemes rely on the existence of efficient spatial indices to perform spatial joins. Abel et al. [Abe95] formalise a spatial semi-join operator, which is a generalisation of the conventional semi-join operator, and present efficient algorithms for performing distributed spatial joins. They present two algorithms based, again, on specialized spatial indices: (1) single dimensional object mappings, commonly known as space filling curves, for point data, and (2) the R tree for region data.

These indices are rarely available in autonomous archives. Furthermore, since their objective is to compute a distributed join over a LAN, they assume that processing costs are relatively "more" than transmission costs. This is not true for federated spatial joins over the Internet, and so their algorithms aren't well suited to our purpose.

## 5. SkyQuery: Architecture and Implementation

SkyQuery is a prototype of an evolving federation of geographically separate astronomy databases on the Internet. It has a web interface that allows astronomers to query spatial data stored in the databases participating in the federation. In particular, SkyQuery evaluates a probabilistic federated spatial join query called the cross match query. In this section, we describe the algorithms and the architecture used in SkyQuery. We begin with a description of the architectural components and their interaction using Web services. Next, we explain the cross match query, and finally, we show how this framework for federation can be used for efficiently evaluating such queries.

### 5.1 Architecture

The SkyQuery architecture is based on the wrapper-mediator architecture (Figure 1), which is common in federated database systems. It consists of three components (1) the Clients, (2) the Portal and (3) the SkyNodes. The Portal is the mediator and the SkyNodes contain the wrappers around databases. These three components coordinate with each other to perform two tasks: allowing databases to join the federation, and evaluating cross match queries.

The Clients are web interfaces (or similar applications) that accept user queries and pass them on to the Portal. The Portal mediates between the Clients and the SkyNodes. It accepts user queries from the Clients and and coordinates with the SkyNodes to evaluate the queries. Primarily, SkyNodes are the heterogeneous, autonomous databases participating in the federation. Each SkyNode also implements services that act as wrappers and hide its DBMS and other platform specific details. This presents a uniform view to the Portal.

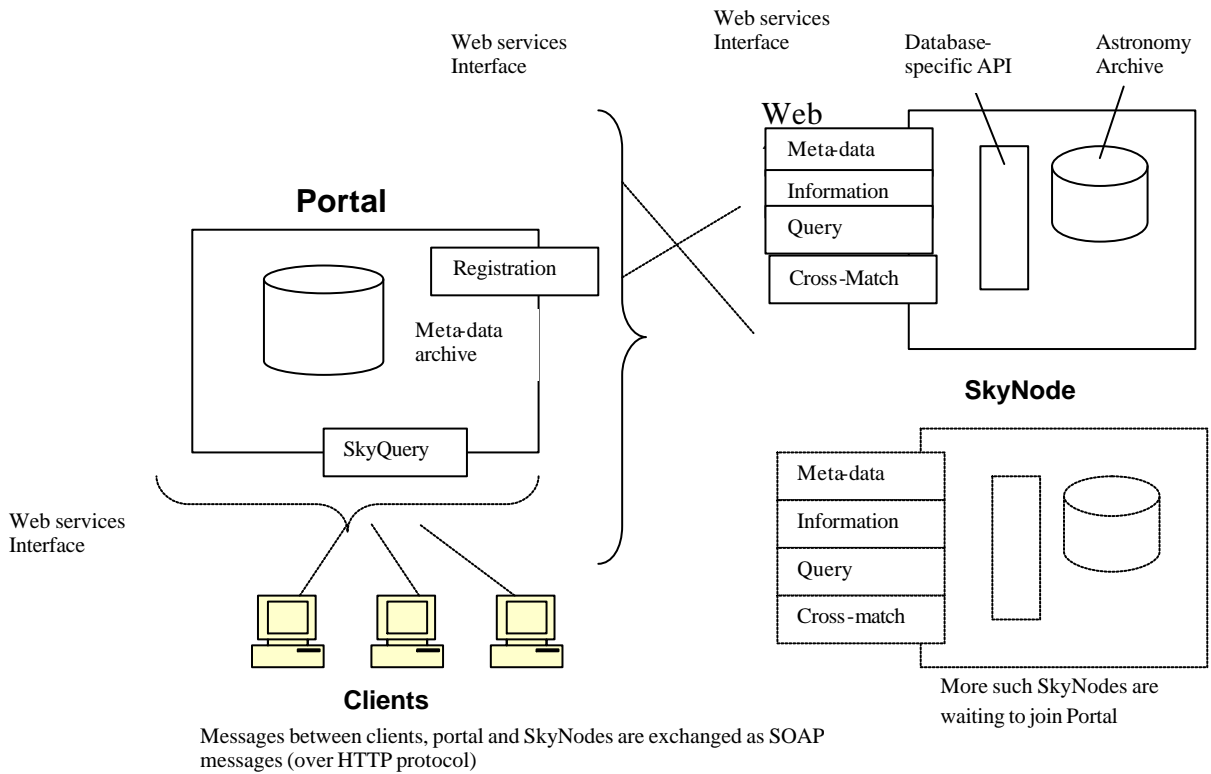

*Figure 1 The SkyQuery Architecture*

The Portal provides two functionalities. First, it implements a *Registration* service that SkyNodes can use to join the federation. As part of registration, the Portal catalogs meta-data-information, which is used in query decomposition. Secondly, it receives SQL-like queries from the Client through its *SkyQuery* service. It decomposes the queries to generate *performance queries* that are used for query optimization. Based on the results of performance queries, the Portal constructs an optimized execution plan, which is, essentially, an ordered set of spatial queries. This order defines the routing of the spatial query set among the participating databases. Many federations, based on the wrapper-mediator architecture, pull results from each database to the Portal. SkyQuery, instead, moves the partial results of spatial queries from one SkyNode to the next along a chain that ends at the Portal. The Portal relays the final result back to the Client.

The SkyNode databases usually have very similar logical schemas. A primary table stores the unique sky position for each astronomical object. Other tables store secondary observations like light intensities in various wavelengths, spectrograms and spectral lines of selected objects, etc. The implementation of the schemas, however, varies, and is an autonomous decision of the organization hosting the database. Apart from this, SkyNode databases have libraries that support range queries. Many astronomy archives support the HTM library [Hie02]; the HTM is a hierarchical spatial index based on spherical triangles. It helps in reducing spatial processing at individual databases. SkyNodes have APIs, which are database specific that provide access to the data and meta-data. In addition, each SkyNode has a wrapper, which implements its specific federation task. This wrapper interfaces with the APIs, as well as the libraries implementing the Web services standard. In essence, the wrapper hides the heterogeneity of the APIs and provides a uniform view to the Portal. This obviates the need for a different module at the Portal to communicate with each SkyNode. (Figure 1).

A SkyNode implements the following four Web services – *Information* service, *Meta-data* service, *Query* service and *Cross match* service. The first three Web services are the minimum set of services needed to provide a basic framework for the federation of astronomy databases. The fourth service performs the actual federated task, which, in our case, is the execution of a cross match query. When a SkyNode wishes to join the SkyQuery federation; it calls the *Registration* service of the Portal. The registration request includes information about services available on the SkyNode. The Portal, in response to this request, calls the *Meta-data* service at the SkyNode. The *Meta-data* service is responsible for providing complete schema information to the Portal, which the Portal catalogs. Once the Portal successfully recognizes a SkyNode, it calls the *Information* service to collect certain astronomy specific constants of that SkyNode such as the object position estimation errors, the name of primary table that stores the position of objects, etc. The *Query* service is a general-purpose database querying service. In our case, it is used by the Portal to answer performance queries. The *Cross match* service executes the cross match query. The next sub-sections explain the cross match query and describe its evaluation using these services.

## 5.2 Cross Match Queries

A cross match query is a SQL-like query with special clauses to specify spatial constraints. The spatial clauses are (1) AREA to specify a spatial range; and (2) XMATCH to specify a probabilistic (fuzzy) spatial join.

The AREA clause is used to specify a circular range in the sky. All objects returned by the query have to lie within this range.

The XMATCH clause is used to match objects in different archives that are observations of the same astronomical body (In this sub-section we use the term "body" to refer to the real astronomical object, and the term "object" to refer to an observation of such a body stored in an archive). Each archive measures the unique position in sky, i.e., *right ascension* and *declination*, for the astronomical bodies it observes. This measured position would be sufficient for exact matches between objects in different archives, except that there is always some unavoidable error in the measurements. The measured position varies slightly, from archive to archive. Astronomers assume that this measured position is a random variable distributed normally around the real position of the body. The standard deviation $\sigma$ of the error is circular and is known for each survey; its value depends on the accuracy of the survey's measuring instruments. Using this knowledge, given a set of objects, one from each archive, one can compute the probability of their being a cross match, i.e., of their referring to the same astronomical body. For a given query, the XMATCH clause specifies the minimum probability threshold for a set of objects to be considered a cross match.

For example, take the following cross match query:

SELECT
    O.object_id, O.right_accession, T.object_id
FROM
    SDSS: Photo_Object O,
    TWOMASS: Photo_Primary T,
FIRST: Primary_Object P WHERE
    AREA (185.0,-0.5,4.5) AND
    XMATCH (O,T,P) <3.5 AND
    O.type= GALAXY AND (O.i_flux - T.i_flux)>2.

In this query, SDSS: Photo_Object, TWOMASS: Photo_Primary, and FIRST: Primary_Object are tables from three different federated SkyNodes, namely, SDSS, TWOMASS, and FIRST. The AREA clause limits the query to a circular range centered at longitude 185.0 and latitude -0.5, and with a radius of 4.5 arc seconds.

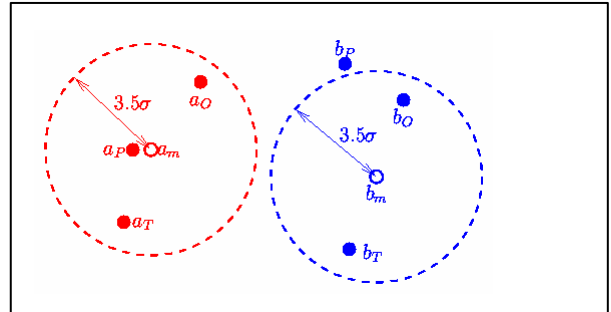

*Figure 2 Different sets of objects are selected based on XMATCH clause specified in user query. The figure shows that set $\{a_O, a_T, a_P\}$ gets selected when user specifies XMATCH(O,P,T) < 3.5 while $\{b_O, b_T, b_P\}$*

The XMATCH clause specifies that only those sets of objects, one each from the three archives, which are within 3.5 standard deviations of their mean position, should be considered. (There is a direct correspondence from this specification to the specification in terms of the minimum probability threshold for a set of objects that was mentioned earlier). For example, in Figure 2, $a_O$, $a_T$, and $a_P$ are the positions for the body $a$ as observed by the archives $O$, $T$, and $P$. Similarly $b_O$, $b_T$, and, $b_P$ are observations for body $b$. Assume that all these observations satisfy the AREA clause. The mean positions for the two sets of observations are $a_m$ and $b_m$, respectively. Note, that the mean positions depend on which observations are being considered for computing the mean, but for the sake of simplicity, we will assume that they are always at the positions shown. The figure also shows, for each body, the range of 3.5 standard deviations from the mean position. All three observations for the object $a$ are within this range. So the set $\{a_O, a_T, a_P\}$ is a cross match. But the set for body b, $\{b_O, b_T, b_P\}$, is not a cross match as $b_P$ is out of range.

The XMATCH clause can also specify the absence of a match. For example, if instead, the clause was XMATCH (O, T, !P) <3.5, it would imply selection of only those sets of objects, one each from the first two archives, that are within 3.5 standard deviations of their mean position, and which don't have a matching object in the last archive within the same error bound. The last archive is called a *drop out*. Note that a drop out specification corresponds to an "exclusive" outer join. In

Figure 2, $\{a_O, a_T, a_P\}$ is not a cross match for this specification, but $\{b_O, b_T\}$ is. Archives specified in XMATCH clause that are not *drop outs* are called *mandatory*.

## 5.3 Query Execution and Optimization

Query execution (see Figure 3) in SkyQuery incurs processing costs at the individual SkyNodes and transmission costs in sending partial results from one SkyNode to the next. We reduce processing costs by using an efficient algorithm for cross match computation at each SkyNode, and we reduce transmission costs by first sending performance queries to estimate the maximum size of the partial result that each SkyNode will transmit.

There are methods that involve extensive calculations for reducing transmission costs, but we use one that is simple, yet effective. Our method follows the basic approach of treating component DBMSs as black boxes, running test queries on them, and finally estimating transmission costs from the results. This approach has attracted considerable attention for its simplicity [Du92, Zhu96]. We call our method the *count star* approach, as our test queries are count queries on the component databases of the federation.

For every user submitted cross match query, the Portal creates performance queries (Step 2 in Figure 3), one for each of the mandatory archives contained in the XMATCH clause. A performance query for a component database is a SQL query constructed to find an upper bound on the number of tuples the database will return

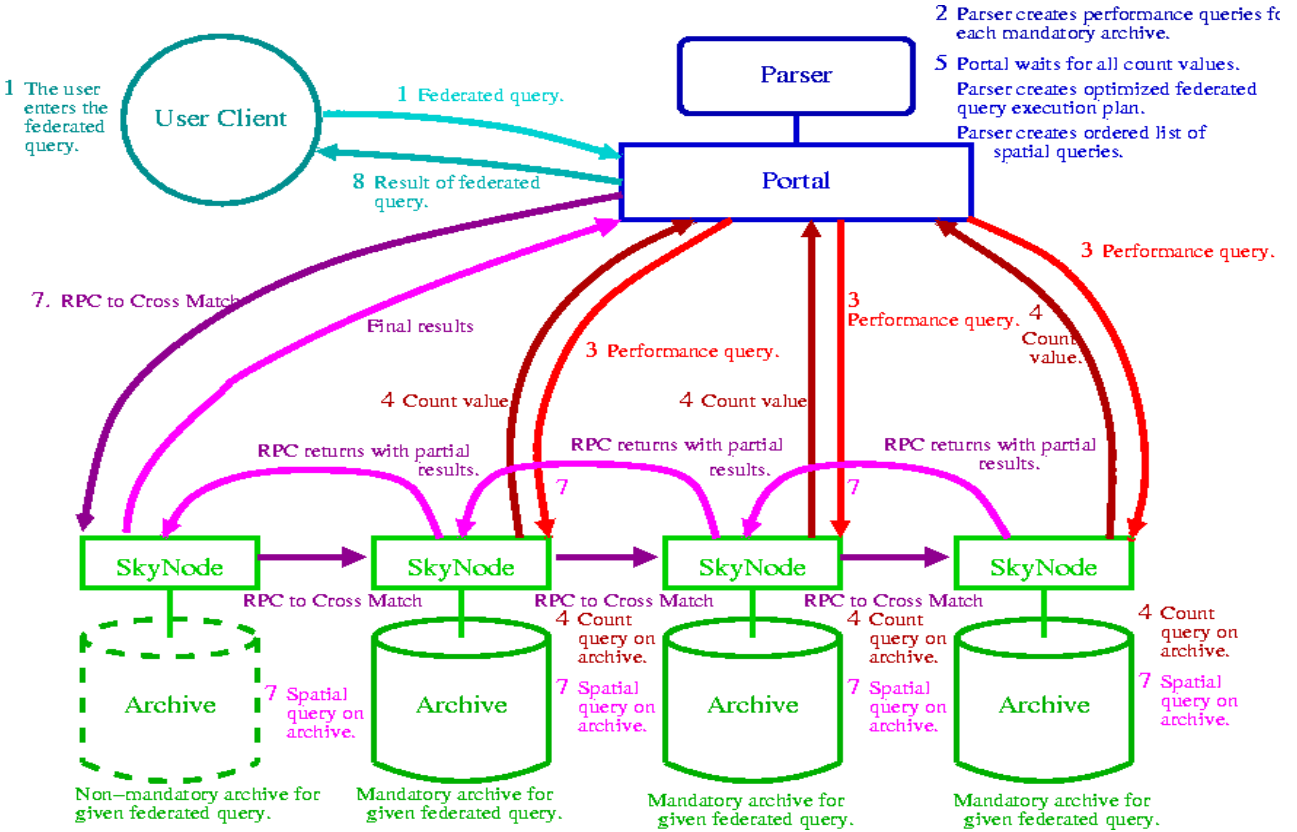

*Figure 3 The Portal and the SkyNodes coordinate to solve a cross match query submitted by a Client. The figure shows the order in which the sample query gets executed.*

during the cross match computation. For example, the cross match query discussed earlier will result in the following three performance queries:

SELECT count(*) FROM SDSS: Photo_Object O
WHERE AREA(185.0,0.5,4,5) AND O.type = GALAXY.

SELECT count(*) FROM TWOMASS: Photo_Primary T
WHERE AREA(185.0,0.5,4,5)
SELECT count(*) FROM FIRST: Primary_Object P

WHERE AREA(185.0,0.5,4,5)

Note, these queries contain clauses that can be evaluated entirely at their SkyNodes. These performance queries are passed as asynchronous SOAP messages to the respective *Query* services of each SkyNode (Step 3 in Figure 3). This will often warm the database cache on each SkyNode with index pages that satisfy the main cross match query, and thus aid in reducing processing time that would be incurred while doing the cross match.

When the messages of performance query results with the number of potentially qualifying tuples arrive at the Portal (Step 4 in Figure 3), it creates a federated query execution plan (Step 5 in Figure 3). (De-serialization of these messages is not an expensive operation as they are single integers.) The federated query execution plan consists of a list of ordered pairs, each containing a query and the URL information of the SkyNode where it would be executed. The list is in *decreasing* order of the count star values returned by the performance queries, with the drop out archives, if any, at the beginning of the list.

To begin cross match query execution, the Portal sends a SOAP RPC to the *Cross match* service of the first SkyNode on the list, and passes the query execution plan as a parameter. This *Cross match* service, in turn, calls the similar service at the second SkyNode on the list, again with the plan as a parameter.

This daisy chaining continues until the last SkyNode on the list, which executes the query meant for it (Step 6 in Figure 3). The query selects rows that satisfy the AREA and other clauses. The query result includes spatial attributes required for computing the cross match. The SkyNode returns this result, as a serialized XML encoded SOAP message to its calling SkyNode. (Step 7 in Figure 3).

Now, at this (second last in the list) SkyNode the encoded results are first de-serialized using a SOAP translator. Next, the results are cast into a database object characteristic to the database that the SkyNode hosts. The *Cross match* service uses the database specific API at that SkyNode to insert the values in the database object into a temporary table. Next, a stored procedure encoding the cross match algorithm (See Section 5.4) uses this temporary table and the primary table at this SkyNode to identify matching objects (with the last SkyNode) that have a potential to satisfy the XMATCH clause. Note, the certainty of the XMATCH clause being satisfied is only known after all the SkyNodes have been consulted. This procedure, in fact, computes an implicit spatial join. (The join is akin to an outer join or an inner join based on whether the archive is a drop out or not). After this, the *Cross match* service executes its own (non-spatial) query from the execution plan and joins - this is an explicit inner join - with the results of the spatial join. The final results are returned to the calling SkyNode. The temporary table is deleted.

The process above is repeated at every SkyNode in the reverse order of the calls. The first SkyNode returns the results to the Portal, which relays it to the client. In this whole process, it is clear that the order based on the count star values will often decrease the network transmission costs.

### 5.4 The Cross Match Algorithm

The AREA clause in a cross match query is implemented using the range search capabilities of the individual archives. We assume that each archive has a reasonably efficient mechanism for implementing a range search. One such mechanism, which is also used by the SkyQuery test databases, is the HTM [Hie02]. This mechanism builds a quad tree on the sky, each node of which corresponds to a spherical triangle. To retrieve objects in a given range, triangles that are entirely within or intersect the range are first computed. All objects in the triangles that are entirely within the range are in the range too. Objects that are in intersecting triangles, however, are again individually tested to check if they are within the range or not. Objects in other triangles are not in range. Similarly, the other clauses, except XMATCH, are implemented using each archive's ability to evaluate regular queries.

The XMATCH clause is implemented by computing a probabilistic spatial join. Let the clause contain $N$ archives. Assume that none of them are dropouts; the extension to include dropouts is straightforward. We first describe, given an $N$-tuple of objects $R = (o_1,...,o_N)$, one from each archive, how to compute the log likelihood of each $o_i$, $1 \le i \le N$, being an observation of the same astronomical body. Let the position for object $o_i$ (from archive $i$) be given by $(x_i, y_i, z_i)$. Let $(x, y, z)$ be the unknown true position of the astronomical body. The log likelihood that the observations in $R$ are of this object is

$$c^2(x,y,z) = -\sum_{i=1}^{N} \frac{1}{2s_i^2}\left[(x-x_i)^2 + (y-y_i)^2 + (z-z_i)^2\right]$$

where $s_i$ is the standard deviation of the error at archive $i$. The best such position $(x, y, z)$ is obtained by minimizing the following chi-squared expression

$$\sum_{i=1}^{N} \frac{1}{2s_i^2}\left[(x-x_i)^2 + (y-y_i)^2 + (z-z_i)^2\right] - \tfrac{1}{2}\mathbf{l}\left[x^2 + y^2 + z^2 - 1\right]$$

where the second term is the Lagrange multiplier that forces $(x, y, z)$ to be a unit vector. This best position is along the direction $(a_x, a_y, a_z)$, and the log likelihood at this position is, $-a + \sqrt{a_x^2 + a_y^2 + a_z^2}$, where

$$a = \sum_{i=1}^{N} \frac{1}{\sigma_i^2},\ a_x = \sum_{i=1}^{N} \frac{x_i}{\sigma_i^2},\ a_y = \sum_{i=1}^{N} \frac{y_i}{\sigma_i^2},\ a_z = \sum_{i=1}^{N} \frac{z_i}{\sigma_i^2}.$$

Only those *N*-tuples satisfy the XMATCH clause that have a log likelihood value greater than the threshold specified in the clause. Let the set of satisfying tuples be *S*. This set is computed in a distributed fashion as each archive is at a different site; the computation is done in *N* steps, one at each archive. At archive *i*, the set $S_i$ consisting of tuples of length *i* that "have the potential" to satisfy the clause is constructed. Let $R_i = (o_1, ..., o_i)$ be one such tuple. For each such $R_i$ the corresponding cumulative values are also computed. Remember that $(a_{i,x}, a_{i,y}, a_{i,z})$ is the direction of the best position $(x_i, y_i, z_i)$ of the astronomical object if $R_i$ is to be a cross match. To compute these values, the archive *i* receives from archive *i*-1 tuples of type $R_{i-1} = (o_1, ..., o_{i-1})$ and the corresponding values $a_{i-1}, a_{i-1,x}, a_{i-1,y}, a_{i-1,z}$. For each $R_{i-1}$, archive *i* retrieves all objects that are close to the current best position $(x_{i-1}, y_{i-1}, z_{i-1})$ (and satisfy the other clauses in the cross match query) using a range search. Each such object is appended to $R_{i-1}$ to get tuples of type $R_i$. The cumulative values for each such $R_i$ are then computed using the information available. Next, the log likelihood of $R_i$ being a cross match is computed, and only if it is larger than the threshold, $R_i$ (and its cumulative values) is sent to the next archive *i*+1.

The first archive just needs to send 1-tuples comprising of objects that satisfy the other clauses in the query to the second archive. The tuples surviving the *N*th archive are the required cross matches. This XMATCH scheme is fully symmetric; the particular order of the archives considered doesn't matter. We do, however, impose an order, using the count star approach to reduce network data transfer costs.

## 6. Conclusion

We want to highlight some of our experiences while developing SkyQuery, talk about extensions, as well as speculate on what it may imply for the future of federations (of astronomy archives) on the Internet.

Since SOAP is an evolving standard, it was difficult to find a fully standard-conforming, well-documented SOAP implementation. The Web service implementation did not require any significant programming overhead as, we feel, would have been required with an alternative technology like CORBA. We faced problems while trying to match large number of objects, though. The XML parser at the SkyNode would run out of memory while parsing SOAP messages of about 10 MB. We worked around by dividing large data sets into smaller chunks. SOAP is considered to be slower than other middleware, like, CORBA, because of the time spent for serialization and de-serialization [XML02].

XML and SOAP are technologies that are still evolving and probably quite far from realizing their peak in terms of the service and the convenience they provide. Yet, we found that they were more than sufficient for building SkyQuery. This makes us want to extend SkyQuery, using XML and SOAP, to perform other complex distributed computing tasks for the Virtual Observatory. We would also like to include a lot more participating archives. That would effectively test the interoperability and the system integration capabilities of Web services. Another extension is to implement transaction processing for exchange of data between astronomy archives, and see how the stateless SOAP handles such complex requirements.

We are still building SkyQuery and extending its functionality. We want to evaluate spatial queries more complex than cross matches. For example, queries on spatial objects that are not just points but objects of different shapes like polygons. They AREA clause can also be extended to specify arbitrary polygons rather than just simple circles. This would imply many more evaluations of spatial joins and an exchange of a larger magnitude of XML data.

SkyQuery shows that Web services are a good medium for publishing data on the Internet. It demonstrates that such archives can be federated with minimal additional software effort and compliance requirements.

## 7. Acknowledgements

The authors thank Jim Gray for inspiring them to write this paper in the first place as well as for the helpful suggestions and discussions at all stages of this project.